# E-NET MODELS FOR DISTRIBUTION, ACCESS AND USE OF RESOURCES IN SECURITY INFORMATION SYSTEMS

**Nikolai Todorov Stoianov, Veselin Tsenov Tselkov**

***Abstract:*** *This paper presents solutions for distribution, access and use of resources in information security systems. The solutions comprise the authors' experience in development and implementation of systems for information security in the Automated Information Systems. The models, the methods and the modus operandi are being explained.*

## 1. INTRODUCTION

Distribution of the resources in security computer systems has an important role for the correct work of system and for protecting the secret. The objective of any distribution of computer resources is to ensure users' access to the right resources in any moment [2,4,7]. A lot of resources are used in a computer system. Some of them are [1,3]:
- files;
- databases;
- sharing devices;
- network devices;
- printers.

The users and the resources can be described with a set of characteristics. Users' rights depend on the characteristic value. According to user's identification time the models for distribution and control are [5]:
- static model;

- dynamic model;
- hybrid model.

These models for distribution enable to define only the interaction user/resource but not user/user which is one of the most frequently used in computer systems.

Another way to distribute computer resources is by security levels and groups. This way of distribution doesn't give way to define the possible interaction between user/user and user/resource [6].

## 2. E-NET MODELS FOR DISTRIBUTION, ACCESS AND USE OF RESOURCES BY GROUPS AND SECURITY LEVELS

In a security computer system let us denote:
- $\{U(i)|\ i=1,2,\ldots,n\}$ is a set of users;
- $\{G(j)|\ j=1,2,\ldots,m\}$ is a set of groups;
- $\{R(k)|\ k=1,2,\ldots,p\}$ is a set of resources;
- $\{L(l)=l|\ l=1,2,\ldots,q\}$ is a set of security levels.

***Principle of interaction in the model:*** *Any group consists of users and resources. Every user and every resource has maximum access level in the group. Any user can interact only with the resources that belong to the same group and simultaneously the user's security level in the group is higher or equal to the resource's level.*

The users of the system are distributed as follows:

For every i (i=1,2,…,n) and for every j (j=1,2,…,m) should be defined Lug(i,j):
- if Lug(i,j)=0, then user U(i) doesn't belong to group G(j);
- if Lug(i,j)=Lu(i,j), where Lu(i,j)$\in\{1,2,\ldots,q\}$ is a number that shows the maximum security level of user

U(i) in group G(j), then user U(i) belongs to group G(j).

The resources of the system are distributed as follows:
For every k (k=1,2,…,p) and for every j (j=1,2,…,m) should be defined Lrg(k,j):
- if Lrg(k,j)=0, then resource R(k) doesn't belong to group G(j);
- if Lrg(k,j)=Lr(k,j), where Lr(k,j)∈{1,2,…,q} is a number that shows the maximum security level of resource R(k) in group G(j), then resource R(k) belong to group G(j).

**Definition 1:** We say that user $U(i_0)$ $i_0 \in \{1,2,…,n\}$ can interact with user $U(i_1)$ $i_1 \in \{1,2,…,n\}$ in group G(j) j∈{1,2,…,m} by resource R(κ) k∈{1,2,…,p} only in case that:
$\min(Lug(i_0,j), Lug(i_1,j)) \geq Lrg(k,j)$.

The functional possibilities for user are:
- request for access (Ident);
- analysis of user rights (CheckAuthorities);
- identification of security level (IdentLevel);
- receiving list of groups (ListGroups);
- selecting group (SelectGroup);
- receiving list of resources (ListResources);
- selecting resource (SelectResource);
- using resource (UseResource);
- saving log file (LogFile);
- exit from the system (Quit).

The E-net model ENGL=<B, Bp, Br, T, F, H, Mo> for distribution, access and use of resource by groups and security levels is shown on fig.1.

B={bp1, br1, br2, b1,..., b9} is the set of model ENGL's places.

T={t1,…, t9} is the set of transitions.

The places and the transitions in the model are in the sense of general places and general transitions.

The relations between places and transitions (functions F and H) are shown on fig.1.

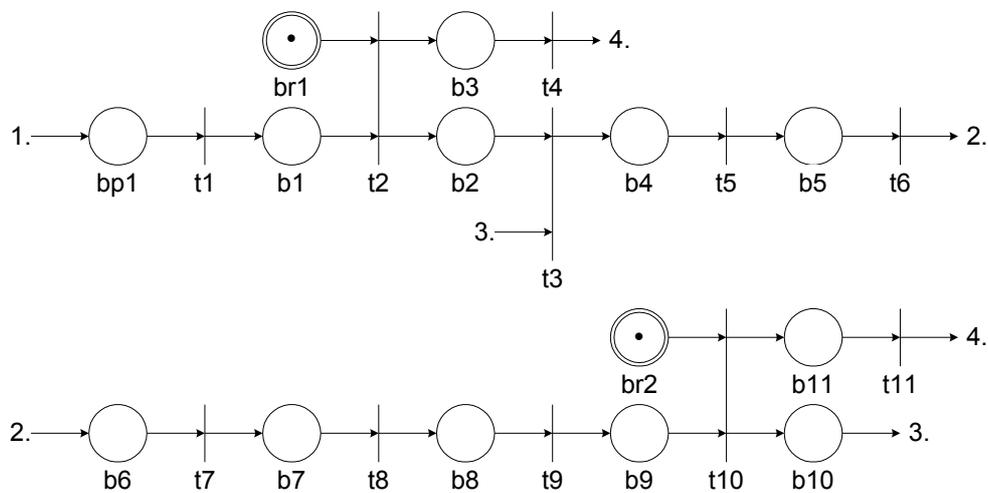

Fig.1 ENGL model for distribution, access and use of resource by groups and security levels

Places of the model ENGL.

Places of the model describe the state and interaction of the users with the system.

Bp={bp1} is the set of peripheral positions and kernel has appeared in bp1 just when a user wants to utilize the system.

Br={br1, br2} is the set of permissive places respectively at transitions t2 and t8.

Transitions of the model ENGL.

Transitions of ENGL simulate:
- t1- request for access (Ident);
- t2 - analysis of user rights (CheckAuthorities);

- t3 - receiving list of groups (ListGroups);
- t4 and t11 - exit from the system (Quit);
- t5 - selecting group (SelectGroup);
- t6 - identification of security level (IdentLevel);
- t7 - receiving list of resources (ListResources);
- t8 - selecting resource (SelectResource);
- t9 - using resource (UseResource);
- t10 - saving log file (LogFile).

Kernels of the model ENGL.

Kernels' descriptions in different model ENGL's places correspond to the input (output) parameters of according transitions.

## 3. E-NET MODELS FOR DISTRIBUTION, ACCESS AND USE OF RESOURCES BY SECURITY LEVELS AND GROUPS

In a security computer system let us denote:
- $\{U(i)| i=1,2,…,n\}$ is a set of users;
- $\{G(j)| j=1,2,…,m\}$ is a set of groups;
- $\{R(k)| k=1,2,…,p\}$ is a set of resources;
- $\{L(l)=l| l=1,2,…,q\}$ is a set of security levels.

***Principle of interaction in the model:*** *Every user and every resource has maximum access level in the group. Any group consists of users and resources. Any user can interact only with these resources which level is lower or equal to the user's level and belong to the same group simultaneously.*

To every user $U(i)$ $i \in \{1,2,…,n\}$ should be added a number $Lu(i)$ $Lu(i) \in \{1,2,…,q\}$ shows the maximum security level.

For every i (i=1,2,..,n) and for every l (l=1,2,…,Lu(i)) should be defined $Ulg(i,l,j)$:
- if $Ulg(i,l,j)=0$, then user $U(i)$ doesn't belong to group $G(j)$ in level $L(l)$;

- if $Ulg(i,l,j)=1$, then user $U(i)$ belongs to group $G(j)$ in level $L(l)$.

To every resource $R(k)$ $k \in \{1,2,\ldots,p\}$ should be added a number $Lr(k)$ $Lr(k) \in \{1,2,\ldots,q\}$ shows the maximum security level.

For every k (i=1,2,..,p) and for every l (l=1,2,…,Lr(k)) should be defined $Rlg(k,l,j)$:
- if $Rlg(k,l,j)=0$, then resource $R(k)$ doesn't belong to group $G(j)$ in level $L(l)$;
- if $Rlg(k,l,j)=1$, then resource $R(k)$ belongs to group $G(j)$ in level $L(l)$;

**Definition 2:** We say that user $U(i_0)$ $i_0 \in \{1,2,\ldots,n\}$ can interact with user $U(i_1)$ $i_1 \in \{1,2,\ldots,n\}$ by resource $R(\kappa)$ $k \in \{1,2,\ldots,p\}$ only in case that:
1. $\min(Lu(i_0), Lu(i_1)) \geq Lr(k)$ and
2. there is $l_0$ ($l_0=1,2,\ldots,Lr(k)$) for which:
   $(Ulg(i_o,l_0,j_0)=1)$ и $(Ulg(i_1,l_0,j_0)=1)$ и $(Rlg(k,l_0,j_0)=1)$.

The functional possibilities for user are:
- request for access (Ident);
- analysis of user rights (CheckAuthorities);
- identification of security level (IdentLevel);
- receiving list of groups (ListGroups);
- selecting group (SelectGroup);
- receiving list of resources (ListResources);
- selecting resource (SelectResource);
- using resource (UseResource);
- saving log file (LogFile);
- exit from the system (Quit).

The E-net model ENLG=<B, Bp, Br, T, F, H, Mo> for distribution, access and use of resource by security levels and groups is shown on fig.2.

B={bp1, br1, br2, b1,..., b9} is the set of model ENLG's places.

T={t1,…, t9} is the set of transitions.

The places and the transitions in the model are in the sense of general places and general transitions.

The relation between places and transitions (functions F and H) are shown in the fig.2.

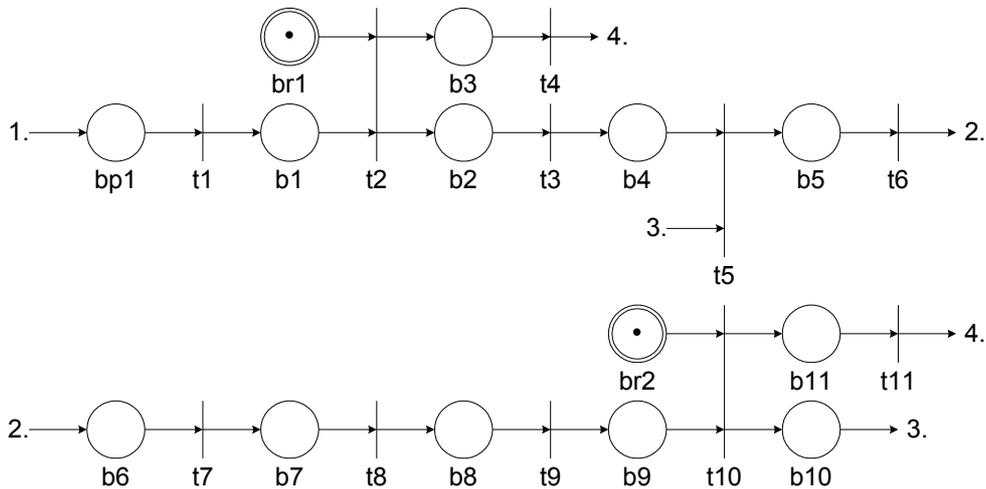

Fig.2 ENLG model for distribution, access and use of resource by security levels and groups

Places of the model ENLG.

Places of the model describe the state and interaction of the users with the system.

Bp={bp1} is the set of peripheral positions and kernel has appeared in bp1 just when a user wants to utilize the system.

Br={br1, br2} is the set of permissive places respectively at transitions t2 and t8.

Transitions of the model ENLG.

Transitions of ENLG simulate:
- t1- request for access (Ident);
- t2 - analysis of user rights (CheckAuthorities);
- t3 - identification of security level (IdentLevel);
- t4 и t11 - exit from the system (Quit);
- t5 - receiving list of groups (ListGroups);
- t6 - selecting group (SelectGroup);
- t7 - receiving list of resources (ListResources);
- t8 - selecting resource (SelectResource);
- t9 - using resource (UseResource);
- t10 - saving log file (LogFile).

Kernels of the model ENLG.
Kernels' descriptions in different model ENLG's places correspond to the input (output) parameters of according transitions.

## 5. SUMMARY

On the bases of the E-net models ENGL and ENLG algorithms and software system for information security is being realized. The suggested models have been approbated in Center for Information Security of Defense Advanced Research Institute.